\documentclass[11pt]{article}
\usepackage{amssymb}
\usepackage{amsmath}
\usepackage[all]{xy}
\usepackage{graphicx}

\title{{\sc Against the Tyranny of\\ `Pure States' in Quantum Theory}}

\author{{\sc C. de Ronde}$^{1,2,3,4}$ and {\sc C. Massri}$^{5,6}$}
\date{}

\usepackage[margin=2 cm]{geometry}

\begin{document}

\bibliographystyle{plain}
\maketitle

\begin{center}
\begin{small}
1. Philosophy Institute Dr. A. Korn, University of Buenos Aires - CONICET\\
2. Center Leo Apostel for Interdisciplinary Studies\\Foundations of the Exact Sciences - Vrije Universiteit Brussel\\
3. Institute of Engineering - National University Arturo Jauretche\\
4. Federal University of Santa Catarina, Brasil\\
5. Institute of Mathematical Investigations Luis A. Santal\'o, UBA - CONICET\\
6. University CAECE
\end{small}
\end{center}

\begin{abstract}
\noindent We argue that the notion of {\it pure sate} within Standard Quantum Mechanics is presently applied within the specialized literature in relation to two mutually inconsistent definitions. While the first ({\it operational purity}) provides a basis-dependent definition which makes reference to the certain prediction of measurement outcomes, the latter ({\it trace-invariant purity}) provides a purely abstract invariant definition which lacks operational content. In this work we derive a theorem which exposes the serious inconsistencies existent within these two incompatible definitions of purity. 
\end{abstract}
\begin{small}

\end{small}

\newtheorem{theo}{Theorem}[section]
\newtheorem{definition}[theo]{Definition}
\newtheorem{lem}[theo]{Lemma}
\newtheorem{met}[theo]{Method}
\newtheorem{prop}[theo]{Proposition}
\newtheorem{coro}[theo]{Corollary}
\newtheorem{exam}[theo]{Example}
\newtheorem{rema}[theo]{Remark}{\hspace*{4mm}}
\newtheorem{example}[theo]{Example}
\newcommand{\proof}{\noindent {\em Proof:\/}{\hspace*{4mm}}}
\newcommand{\qed}{\hfill$\Box$}
\newcommand{\ninv}{\mathord{\sim}} %involutive negation
\newtheorem{postulate}[theo]{Postulate}

\section{Pure States in Quantum Mechanics}

Today, the notion of {\it pure state} plays an essential role within the ongoing research that takes place in the context of what is considered to be the ``Standard'' version of Quantum Mechanics (QM).\footnote{Even though there are many interpretations of QM, some of which change the mathematical formalism of the theory (e.g., Bohm and GRW), there is a ``Standard'' version of QM which is taught in Universities all around the world.} The kernel role of {\it pure state} has also affected the fields {\it Quantum Foundations} and {\it Quantum Information} through its centrality regarding the definition of quantum entanglement. This notion was introduced during the axiomatic formulation of the theory in the 1930s, and has become increasingly dominant establishing also an ontological primacy over the so called {\it mixed states}. As explained by David Mermin \cite[p. 758]{Mermin98b}: ``[P]eople distinguish between pure and mixed states. It is often said that a system is in a pure state if {\it we} have maximum {\it knowledge} of the system, while it is in a mixed state if {\it our knowledge} of the system is incomplete.'' The explicit reference to ``our knowledge'' is strictly related to the widespread ---specially, in the field of quantum information--- operational definition of a {\it pure state} in terms of certain predictions. This definition rests in what is known as a {\it maximal test}: `If a quantum system is prepared in such way that one can obtain a maximal test yielding with certainty (probability = 1) a particular outcome, then it is said that the quantum system is in a \emph{pure state}.' In turn, the notion of {\it maximal test} allows to interpret a quantum observable as being an {\it actual property} ---i.e., a property that will yield the answer {\it yes} when being measured (see for discussion \cite{BeltramettiCassinelli81, FLSG13, Jauch68, Mittelstaedt78, Smets05, Piron76}). It is then stated that the pure state of a quantum system is described by a unit vector in a Hilbert space,  $\mathcal{H}$, which in Dirac's notation is written as $|\psi \rangle$. Depending on the basis, a pure state in $\mathcal{H}$ is also represented by a superposition of states: 
$$|\psi\rangle=\sum a_i|\varphi\rangle_i$$ 
\noindent As it is well known, such superposed states yield uncertain results. Indeterminate or potential properties pertaining to superposed states might or might not become actualized in a future instant of time. These properties are regarded in the orthodox literature as being {\it uncertain} and thus, cannot be considered as elements of physical reality (in the EPR sense \cite{EPR}) nor interpreted in terms of ignorance. Asher Peres explains this important point in the following manner: 
\begin{quotation}
\noindent {\small ``According to quantum theory, we have a choice between different, mutually incompatible tests. For example, we may orient the Stern-Gerlach magnet in any direction we please. Why then is such a Stern-Gerlach test called complete? The reason can be stated as the following postulate:

\smallskip

\noindent {\bf A. Statistical determinism.} {\it If a quantum system is prepared in such a way that it certainly yields a predictable outcome in a specified maximal test, the various outcomes of {\bf any} other test have definite {\bf probabilities.} In particular, these probabilities do not depend on the details of the procedure used for preparing the quantum system, so that it yields a specific outcome in the given maximal test. A system prepared in such a way is said to be in a {\bf pure} state.}'' \cite[p. 66]{Peres02}}
\end{quotation}

It is also accepted in Standard QM that the notion of pure state can be extended to density operators. This is an essential step that has been worked out by mathematical physicists who regard the previous definition as ``too limited'' (see \cite[p. 419]{Hall13}), specially when considering systems and subsystems. Let $\mathcal{H}$ be a Hilbert space, a density operator $\rho$  (i.e., a positive trace class  operator with trace 1) is called a \emph{state}. Being positive and self-adjoint, the eigenvalues of $\rho$ are non-negative and real and it is possible to diagonalize it. Pure states are represented by rank 1 matrices which in their diagonal form will be given by $(1,0,\ldots,0)$ ---where the place occupied by the `1' is unimportant. In this case,  $\rho$ is equal to  $vv^{\dag}$ for some normalized vector $v\in\mathcal{H}$. If the rank of $\rho$ is grater than 1 (or equivalently if $\mbox{Tr}(\rho^2)<1$), the state is called  a \emph{mixed state}; or in short, a {\it mixture}. For example, the vector  $\alpha|0\rangle+\beta|1\rangle$, $|\alpha|^2+|\beta|^2=1$, gives the following density matrix:  
\[
\rho=\begin{pmatrix}
|\alpha|^2 & \alpha\overline{\beta}\\
\overline{\alpha}\beta&|\beta|^2
\end{pmatrix}
\]
Note that, if $\rho$ is a pure state (i.e., $\mbox{Tr}(\rho^2)=1$), there always exists a basis in which the matrix can be diagonalized as:
\[
\rho_{pure}=
\begin{pmatrix}
1 &0\\
0&0
\end{pmatrix}
\]

\noindent Unlike the case of pure states, mixtures cannot be represented as a unit vector, $|\psi \rangle$. Instead, mixed states are conceived as mixtures of pure states and represented as their  convex sums:  
$$\rho_{mix} = \sum_{i} p_{i} \  \rho_{i}^{pure} =  \sum_{i} p_i \ | \Psi_i \rangle \langle\Psi_i | $$
Thus, while pure states seem to refer to the existence of an observable which, if measured, will be obtained with {\it certainty} (probability equal to 1), mixed states do not. There will exist no single context of measurement (no basis) for which a mixed state will predict with certainty a {\it yes-no} answer for a specific observable. Indeed, if $\mbox{Tr}(\rho^2)<1$, then whatever basis we choose to write $\rho$, the value $\mbox{Tr}(\rho^2)$ will remain unaltered. Therefore, the matrix of $\rho$ in that basis cannot represent a pure state since for pure states we have $\mbox{Tr}(\rho^2)=1$. Mixtures are interpreted in terms of ignorance regarding the pure state in which the quantum system really is.\footnote{The reference to `mixtures' ---as contra-posed to `pure states'--- is extremely problematic for it erases the fundamental distinction between {\it quantum mixture} and {\it classical mixture}; a well known distinction in the specialized literature which Bernard d'Espagnat termed {\it proper} and {\it improper} (see \cite[chap. 6]{d'Espagnat76}).} As remarked by Nancy Cartwright in \cite{Cartwright72}: ``The ignorance interpretation asserts that each member of the collection is in one of the pure states in the sum ---it is only our ignorance which prevents us from telling the right pure state for any specific member.'' Thus, contrary to the case of pure states, when considering mixed states, all observables are {\it uncertain}; they all possess a probability which pertains to the open interval $(0,1)$. As stressed by Cartwright: ``The ignorance interpretation is the orthodox interpretation for mixtures, and should not be confused with the ignorance interpretation for superpositions, which has been largely abandoned.'' As an example of a mixed state (i.e., $\mbox{Tr}(\rho^2)<1$) we can consider the following diagonal matrix,
\[
\rho_{mixed}=\begin{pmatrix}
\frac{1}{2} &0\\
0&\frac{1}{2}
\end{pmatrix}
\]

\noindent According to orthodoxy this mixed state provides {\it minimal knowledge} about the pure state the quantum system is presupposed to be in. Thus, it is impossible to know ---in any bases--- which pure state will be obtained after a measurement is performed. Let us remark that this definition of pure state is also essential in the field of quantum information due to its connection to the measure of entanglement through the {\it von Neumann entropy} (see \cite{BZ06}). As remarked by Brian Hall: ``we may define the {\it von Neumann entropy} $S(\rho)$ of a density matrix $\rho$ by $S(\rho) = \mbox{Tr}(- \rho \log \rho)$ where $\rho \log \rho$ is defined by the functional calculus. (Since $\lim_{\lambda\to 0^+}\lambda\log\lambda = 0$, we interpret $0\log0$ as being $0$). Since the eigenvalues of $\rho$ are all between $0$ and $1$, we see that $- \rho \log \rho$ is a non-negative self-adjoint operator, which has a well-defined trace, which may have the value $+\infty$. According to Exercise 8, a density matrix $\rho$ is a pure state if and only if $S(\rho) = 0$.''

\smallskip 

Up to now, we have just given a summary of the orthodox way in which pure states and mixtures are commonly presented in Standard QM. In the present work we attempt to clearly distinguish between these two different definitions of the notion of `pure state' which are indistinctively applied in the specialized literature. While both definitions are assumed to be equivalent, we will show that this is clearly not the case. More specifically, we will expose a series of inconsistencies between these two different definitions, one contextual and another non-contextual, one possessing a clear operational content and the other purely abstract. In order to understand how these inconsistencies ---present within the definition(s) of pure state--- were introduced within Standard QM, we will go back, in section 2, to the original meaning of {\it state} in classical physics. We will then continue to address, in section 3, the re-definition of the notion of state within Paul Dirac's 1930 formulation of the theory of quanta. In section 4 we will be then ready to expose the inconsistencies we are referring to. In particular we will derive a theorem which exposes the problem.

\section{The Notion of `State' in Classical Physics}

As remarked by Arthur Fine \cite[p. 94]{Schlosshauer11}: ``Wolfgang Pauli thought that using the word `state' ({\it Zustand}) in QM was not a good idea, since it conveyed misleading expectations from classical dynamics.'' The notions of {\it system} and {\it state} have always played an essential role in all classical physical theories going from classical mechanics and wave mechanics to electromagnetism and classical statistical mechanics. The way in which these notions were developed is intrinsically related to the role played by two kernel notions that appeared with modern physics, namely, {\it invariance} and {\it objectivity}. In this section, before entering the more specific debate about quantum states and purity, we would like to provide an analysis of the previous understanding of {\it physical state} ---before QM. The notion of {\it system} is essential to classical physics, it comprises the formal-conceptual {\it moment of unity} of which the representation of the physical theory talks about. Each physical theory has constructed its own specific notion of what a system {\it is}. While in classical mechanics the system is given by the notion of `particle' or `rigid body' (i.e., a compound of particles), in wave mechanics its role is played by the notion of `wave' and in electromagnetism the main notion is that of `field'. Every theory has its own mathematical formalism and its own specific notions at play. In all theories, the systems (i.e., particles, fields or waves) provide the basic reference to a specific representation of a real state of affairs which evolves in time. This evolution described through the interaction between systems is computed in mathematical terms via an equation of motion (or evolution). A kernel presupposition in modern physics which helps to build up this type of representation is that the state of affairs described by the theory must be completely {\it detached} from any reference frame or subject. The physical notion of {\it state} has the main purpose to provide the specific characteristics of a system, in a specific given situation or actual state of affairs. For example, the particle 1 has position $p_1$ and velocity  $v_1$ while particle 2 has position $p_2$ and velocity  $v_2$, etc. The specification of the state of each system the theory talks about is always {\it relative} to another essential notion in physics ---one which every student of physics learns in the first class she attends---, namely, that of {\it reference frame}. Students are taught that it is only when considering a frame of reference that the notion of state becomes applicable. In physics, the specific values of the properties of a system (i.e., the state) have meaning only when considered as relative to a specific viewpoint. Thus, without a reference frame there can be no consistent definition of a state. At this point, the notion of {\it physical invariance} becomes essential in order to secure the consistency of the different reference-frame dependent representations of the state of affairs being addressed. It is important to stress that this notion is not purely mathematical (see for discussion \cite{Scholz18}). Mathematical invariance is a very general abstract concept which can be defined in the following manner. Let $G$ be a group acting on spaces $V$ and $W$. A function $f:V\to W$ is  called \emph{invariant} if $f$ maps orbits to orbits, that is,  if $f(g\cdot v)=g\cdot f(v)$ for all $g\in G$ and $v\in V$. Let us give an example of an invariant quantity ($W=\mathbb{C}$). Consider the space of matrices and let $G$ be the group of invertible matrices acting by conjugation. Then, the trace (or the determinant) is invariant, $\mbox{Tr}(\rho)=\mbox{Tr}(U\rho U^{\dag})$ (where $U$ is a unitary matrix). Something which distinguishes the physical notion of invariance is that it is essentially operational, it requires a link with experience. In fact, this was the essential point raised by Albert Einstein in his criticism about the notion of {\it simultaneity}. 
\begin{quotation}
\noindent {\small ``The concept does not exist for the physicist until he has the possibility of discovering whether or not it is fulfilled in an actual case. We thus require a definition of simultaneity such that this definition supplies us with the method by means of which, in the present case, he can decide by experiment whether or not both the lightning strokes occurred simultaneously. As long as this requirement is not satisfied, I allow myself to be deceived as a physicist (and of course the same applies if I am not a physicist), when I imagine that I am able to attach a meaning to the statement of simultaneity. (I would ask the reader not to proceed farther until he is fully convinced on this point.)'' \cite[p. 26]{Einstein20}}
\end{quotation} 
The essential role played by operational invariance is that it allows to consider experience from different reference frames (and observers) in a consistent manner as making reference to the same state of affairs. In turn, as Einstein constantly remarked, this makes possible to address the (real) existence of a state of affairs as {\it detached} from particular observations and reference frames.\footnote{In more general terms, as discussed in \cite{deRondeMassri18a}, it is exactly this formal aspect which allows us to talk in terms of an {\it Actual State of Affairs (ASA)} that evolves in time; i.e., a dynamical description in terms of the variation of (objective) definite valued observables (or `dynamical properties') independent of the (subjective choice of the) perspective (or reference frame) from which they are being observed. Even in relativity theory, due to the Lorentz transformations, one can still consider `events' as the building blocks of physical reality.} 

In physics, the meaning of invariance deals with physical quantities which not only have the same value for any reference frame, but ---as important--- have also a consistent translation of dynamical values. In general, the transformations that allow us to consider the physical magnitudes from different frames of reference have the property of forming a {\it group} (see \cite{Weyl38, Weyl52}). In the case of classical mechanics we have the Galilei transformations which keep space and time apart, while in relativity theory we have the Lorentz transformations which introduce an intimate connection between space and time coordinates. Of course, restricting ourselves to physical magnitudes that remain always the same, independently of the reference frame, does not provide a dynamical picture. Such a description only provides a static table of data. This is of course completely uninteresting for physics, which always attempts to represent, not only a given state of affairs in a given instant of time but ---far more importantly--- how that state of affairs {\it changes}. Thus, that which matters the most for a physical description is the {\it invariant variations} of physical magnitudes, that is, the dynamical magnitudes which vary but can be still considered as being {\it the same} (e.g., position, velocity, momentum, etc.). {\it The difference within the identity.} Even though the values of physical magnitudes might vary from one reference frame to the other, in all classical theories ---including relativity theory--- there exist a {\it consistent translation} between the values of magnitudes of the different reference frames secured by the transformation laws. The position of a rabbit running through a field and observed by a distant passenger of a high speed train (the reference frame $S$) can be translated to the position of that same rabbit now considered from the perspective of another passenger waiting on the platform of the station (the reference frame $S'$). This relation (of the values between $S$ and $S'$) is provided via the transformation laws. Such transformations include not only the dynamics of the observables but also the dynamics between the different observers (see for a detailed analysis \cite{deRondeMassri16}). The fact that the values of position, momentum, etc., can be consistently translated from one reference frame to the other allows us to assume that these physical observables bear an objective real existence completely independent of the specific choice of the reference frame pertaining to each observer. It is this consistency within translation which allows the physicist to claim that: the rabbit has a set of dynamical properties (position, a momentum, etc.) independently of any of the subjects observing it ---in the train, on the platform or anywhere else. Thus, it is {\it operational invariance}, namely, the invariance with respect to the values of measurable properties, which interests the physicist and allows her to constitute what is to be considered {\it the same} irrespectively of the reference frame or basis (i.e., the perspective) from which it is being represented or ---even--- experienced. Invariants capture the objective (non-contextual) content of a theory by providing a consistent translation between different {\it reference frames}. And in this way, reference frames as well as (empirical) subjects become superfluous to the physical representation. 

%The same reasoning can be applied to coordinate transformations in the phase space $\Gamma$. If we consider a set of observables in a coordinate system, $S$, and perform a transformation of coordinates (e.g., a rotation) to a new system, $S'$, then the values of the dynamical properties will be also consistently translated from the system $S$ to the system $S'$. Such consistency, which is again secured by the transformation, is the {\it objectivity condition} which allows us to consider the observables as preexistent ---and independent--- to the choice of the coordinate system (i.e., the mathematical representation from which we choose to describe our system). It is in this way, that mathematical invariance allows us to detach the empirical subject (i.e., the particular agent performing the experiment) from the objective theoretical representation of physical reality.

In QM the reference frame is given by the mathematical notion of basis and the group that allow us to go from one basis to another is the unitary group. It might be noted that from a conceptual perspective, a mathematical frame of reference (or basis) can be linked to a particular perspective of analysis or an experimental situation (see \cite{BeltramettiCassinelli81}). Something which in QM is also known as {\it context}. As we shall discuss in the following section, during the 20th Century the meaning and content of notion of physical state was radically transformed within quantum theory. Unlike in the classical case we have just discussed the notion of {\it quantum state} would, at least partly, become basis dependent or contextual.

\section{The Contextual Definition of `Quantum State'}

Since Paul Dirac's textbook formulation there has been a deep problem in order to address the meaning of {\it states} and {\it systems} in QM. Even today, there still exists within the specialized literature a deep and widespread disorientation related to the meaning of these notions which are commonly applied in inconsistent manners, sometimes referring to abstract mathematical properties while others to ``small particles'' (see \cite[Chap. 4]{Schlosshauer11}). We believe that the reason behind this situation is intrinsically related to the inadequate interpretation, introduced by Dirac, of the the orthodox mathematical vectorial formulation of QM in terms of these notions. 

As we have just seen in the previous section, in modern physics a {\it state} made reference to the specific features of an object in a given physical situation. The essential point is that the different {\it reference frame} dependent representations of the same state of the system could be related through a group of transformations. Galilean transformation in classical physics and Lorentz transformations in Relativity theory. It is the consistency secured by the transformations which allowed to talk of the same state of affairs regardless of the chosen reference frame. However, in his 1930 book {\it The Principles of Quantum Theory}, Dirac related each {\it state} with a single observation, mathematically represented by ---what he himself called--- a {\it ket} vector. As a consequence, it made perfect sense to argue that {\it the same system} could be represented in terms of {\it different states}, depending on the specific basis: 
\begin{quotation}
\noindent {\small ``[E]ach state of a dynamical system at a particular time corresponds to a ket vector, the correspondence being such that if a state results from the superposition of certain other states, its corresponding ket vector is expressible linearity in terms of the corresponding ket vectors of the other states, and conversely. Thus the state $R$ results from a superposition of the states $A$ and $B$ when the corresponding ket vectors are connected by $ | R \rangle\  = c_1 | A \rangle\ + c_2 | B \rangle\ $.'' \cite[p. 16]{Dirac74}}
\end{quotation}
As it becomes clear, depending on the chosen basis (or reference frame), the representation of the system was given in terms of different states. Bases did not make anymore reference to {\it the same} state of the object under study, but instead to {\it different states}. Each reference frame (or basis) made thus reference to a sub-set of possible states of the quantum system.

This radical transformation in the meaning of (quantum) state is related in the literature to the contextual nature of the theory of quanta, something that Niels Bohr would make even more explicit in his famous reply to the EPR paper \cite{Bohr35}. Thus, the classical presupposition of physics according to which a system can only have a single state at a time was abandoned as a prejudice of the past. The quantum state, or superposition of states, was determined by the choice of the basis regardless of any invariant theoretical condition between the different basis. Dirac simply blamed his controversial definition of quantum states on QM itself and the limits it imposed on  the very possibilities of representation. Of course, this was not considered by Dirac \cite{Dirac74} as an essential problem for he had already ``remarked that the main object of physical science is not the provision of pictures, but the formulation of laws governing phenomena and the application of these laws to the discovery of phenomena. If a picture exists, so much the better; but whether a picture exists of not is a matter of only secondary importance.'' As he made the point, it is always ``important to remember that science is concerned only with observable things''. 

What Dirac did not seem to recognize was that his interpretation implied a much deeper commitment, one which precluded the possibility for QM of an invariant operational discourse (section 2). The abstract invariance of unit vectors with respect to the trace seemed enough to Dirac to restore the reference to quantum systems. During the 1940s and 1950s Dirac's state-vectorial formulation was established in textbook QM through the essential introduction of the notion of {\it pure state}. A notion which made even more explicit the double reference, on the one hand, to the abstract trace-invariance of a unit vector, and on the other, to the non-invariant certain prediction of a single observable.\footnote{Today, the reference to states has become explicitly instrumental. As pointed out by Chris Timpson \cite{Timpson10}, for many contemporary researchers, the quantum state does not represent ``how things are in an external, objective world, it merely represents what information one has. Mermin (2001), Peierls (1991), Wheeler (1990) and Zeilinger (1999) have all endorsed this kind of view. Hartle (1968, p. 709) provides an excellent summary: {\it `The state is not an objective property of an individual system but is that information, obtained from a knowledge of how a system was prepared, which can be used for making predictions about future measurements'.}''}  Dirac's redefinition of the notion of state, and the subsequent introduction of the notion of {\it pure state} within the orthodox literature can be also understood as intrinsically related to the double role played by the notion of {\it actuality} within QM. Indeed, actuality has two different ---not necessarily compatible--- meanings and uses which have been systematically confused and scrambled. Firstly, there is an empiricist (or subjectivist) understanding of actuality relative to the {\it hic et nunc} experience of an individual agent. According to Bas van Fraassen \cite[p. 197]{VF80}, one of today's most prominent empiricists: ``the only believe involved in accepting a scientific theory is belief that it is empirically adequate: all that is {\it both} actual {\it and} observable finds a place in some model of the theory. So far as empirical adequacy is concerned, the theory would be just as good if there existed nothing at all that was either unobservable or not actual. Acceptance of the theory does not commit us to belief in the reality of either sort of thing.'' This first meaning of actuality can be resumed in the following manner:
\begin{definition}[Empiricist Actuality] Actuality as making reference to the here and now empirical observations made by individual subjects (or agents).
\end{definition}
Secondly, actuality is also ---implicitly--- understood in metaphysical terms as characterizing a {\it mode of existence} ---completely independent of observations. In the XVII Century, within the Newtonian mechanical description of the world, any indetermination ---related in the Aristotelian scheme to the potential realm of being--- was erased from the physical representation of reality. Within classical mechanics, every physical system could be described exclusively by means of its actual properties. As remarked by Dennis Dieks: 
\begin{quote}
{\small ``In classical physics the most fundamental description of a physical system (a point in phase space) reflects only the actual, and nothing that is merely possible. It is true that sometimes states involving probabilities occur in classical physics: think of the probability distributions in statistical mechanics. But the occurrence of possibilities in such cases merely reflects our ignorance about what is actual. The statistical states do not correspond to features of the actual system (unlike the case of the quantum mechanical superpositions), but quantify our lack of knowledge of those actual features.'' \cite[p. 124]{Dieks10}}
\end{quote}
This second understanding of actuality which can be defined without any reference whatsoever to observability is of course purely formal and metaphysical. As discussed in detail in \cite{deRondeMassri18a}, an {\it Actual State of Affairs} (ASA) can be defined as a closed system considered in terms of a set of actual (definite valued) properties which can be thought as a map from the set of properties to the $\{0,1\}$. Specifically,  an ASA is a function $\Psi: \mathcal{G}\rightarrow\{0,1\}$ from the set of properties to $\{0,1\}$ satisfying certain compatibility conditions. We say that the property $P\in \mathcal{G}$ is \emph{true} if $\Psi(P)=1$ and  $P\in \mathcal{G}$ is \emph{false} if $\Psi(P)=0$. The evolution of an ASA is formalized by the fact that the morphism $f$  satisfies $\Phi f=\Psi$. Diagrammatically, 
\[
\xymatrix{
\mathcal{G}_{t_1}\ar[dr]_{\Psi}\ar[rr]^f&&\mathcal{G}_{t_2}\ar[dl]^{\Phi}\\
&\{0,1\}
}
\]
Then, given that $\Phi(f(P))=\Psi(P)$, the truth of $P\in \mathcal{G}_{t_1}$ is equivalent to the truth of $f(P)\in \mathcal{G}_{t_2}$. This formalization comprises the idea that the properties of a system remain existent through the evolution of the system. The model allows then to claim that the truth or falsity of a property is independent of particular observations. Or in other words, binary-valuations are a formal way to capture the classical actualist (metaphysical) representation of physics according to which the properties of objects preexist to their measurement. 
\begin{definition}[Metaphysical Actuality] Actuality as making reference to a mode of existence defined in terms of definite binary valuedness of properties which evolve according to the metaphysical representation ---completely independently of subjects and their observations or measurements.
\end{definition}

Maybe the best exposure of this scrambling can be found within the famous definition of {\it element of physical reality} presented in the famous 1935 paper by Einstein, Podolsky and Rosen \cite{EPR}.

\smallskip
\smallskip

\noindent {\bf Element of Physical Reality:} {\it If, without in any way disturbing a system, we can predict with certainty (i.e., with probability equal to unity) the value of a physical quantity, then there exists an element of reality corresponding to that quantity.}

\smallskip
\smallskip

\noindent As remarked by Diederik Aerts and Massimiliano Sassoli de Bianchi \cite[p. 20]{AertsSassoli17}: ``An element of reality is a state of prediction: a property of an entity that we know is actual, in the sense that, should we decide to observe it (i.e., to test its actuality), the outcome of the observation would be certainly successful.'' It is in this way that the relation between observation and reality is pasted together. Reality is not anymore defined in terms of a theoretical representation, but instead in terms of {\it certain} observability. The pasting is also exposed by the dualistic reference to both {\it properties} and {\it observables} in the context of QM. This shift marks the subtle transmutation form a metaphysical discourse to a subjectivist one. It is interesting to note that the notion of {\it pure state} in QM has an analogous role to the one played by actuality within the present widespread empirical-positivist understanding of physical theories. Just like the notion of actuality has a double reference, on the one hand to a metaphysical mode of existence (actual property) and on the other to empirical observation (actual observable), the notion of pure state scrambles a specific type of contextual measurement in which {\it certainty} is restored with a non-contextual purely abstract mathematical definition which lacks any clear operational content. As we shall see in the following section, this tension ---threatening inconsistency--- found within the definition(s) of `pure state' is not only present at the conceptual level of analysis, it is also ---maybe more importantly--- also present within its formal definition itself.

\section{The Inconsistent Definition(s) of `Pure State'}

An exposure of the tension found within the already mentioned incompatible reference to the notion of `actuality' is also present in QM within the definition(s) of {\it pure state}. Let us take as an example the famous book by Michael Nielsen and Isaac Chuang, {\it Quantum Computation and Quantum Information}, where they introduce the notion of {\it pure state} in the following orthodox manner: 
\begin{quotation}
\noindent {\small ``A quantum system whose state $|\Psi\rangle$ is known exactly is said to be in a pure state. In this case the density operator is simply $\rho =  |\Psi\rangle \langle \Psi|$. Otherwise, $\rho$ is in a mixed state; it is said to be a mixture of the different pure states in the ensemble for $\rho$. In the exercises you will be asked to demonstrate a simple criterion for determining whether a state is pure or mixed: a pure state satisfies $\mbox{Tr}(\rho^2)=1$, while a mixed state satisfies $\mbox{Tr}(\rho^2)<1$.'' \cite[p. 100]{NielsenChuang00}}
\end{quotation}
The attentive reader might have already observed, there exists in this passage an inconsistent reference to two different definitions of purity. While in the first phrase there is an operational reference in terms of the ``exact knowledge'' of the state; in the last phrase, the reference is made to the trace-invariance of abstract vectors. The idea that these two definitions are somehow equivalent might be regarded as one of the fundamental widespread confusions present within the contemporary  Standard understanding of QM. Let us analyze these two different definitions in some detail. 

The operational (contextual) definition of pure state to which Nielsen and Chaung make reference in the first phrase rests on a specific type of measurement called {\it maximal test} ---which we already mentioned in the introduction. A test is considered maximal in the case we obtain with {\it certainty} (probability = 1) the observable in question: if we measure the state $|\psi \rangle$ (in its correspondent basis) we are certain that we will obtain the measurement result related to this state. Consequently, we obtain ``complete'' or ``exact knowledge'' with respect to the future prediction of an outcome. This definition rests on the explicit reference to the {\it preferred basis} (or context) in which the vector in Dirac's notation can be written as a single term, namely, as $|\psi \rangle$. We say that the definition is `contextual' because it is explicitly dependent on one particular basis ---between the many possible ones. It is only when we arrange the experimental set up in this {\it preferred basis} that  we will be able to predict a {\it certain} (probability = 1) result (see \cite{BeltramettiCassinelli81, FLSG13, Jauch68, Smets05}). However, if we change the basis, the state $|\psi \rangle$ will be written as a superposition of other different states, $\sum_{i} c_i \ | \Phi_i \rangle$, and in such case, there will be no certain or complete knowledge of any observable (related to  $|\Phi_i \rangle$). We will not be certain of which outcome will be obtained. In other words, the {\it operational purity} of a quantum system (i.e., the certain knowledge of a future measurement) can be found in only one particular reference frame (or basis). Exactly the same actualist intuition appears in the case of density operators where the state $\rho$ is pure only if there exists a basis in which the matrix can be diagonalized as:
 \[
\rho_{pure}=
\begin{pmatrix}
1 &0&\ldots&0\\
0&0&\\
\vdots&&\ddots\\
0&0&\ldots&0
\end{pmatrix}
\]
\noindent This contextual definition has the purpose to secure the existence of one observable which will be {\it certain}, and consequently {\it actual}, if measured. 
\begin{definition}[Operational Purity] Given a quantum system in the state $|\psi \rangle$,  there exists an experimental situation (or context) in which the test of it will yield with certainty (probability = 1) its related outcome. 
\end{definition}
\noindent In short, as remarked in \cite[p. 25]{BZ06}: ``The pure states are those for which the outcome is certain, so that one of the [probabilities] $p_i$ is equal to one.'' As we discussed in the previous sections, this definition is highly problematic for an understanding which respects the basic democratic standpoint of physics according to which a physical concept must be independent of the choice of a particular reference frame. But since the above definition of {\it operational purity} makes explicit reference to a {\it preferred basis}, the concept appears as related to only one single reference frame. Or in other words, the operational content of purity (i.e., the certain prediction of a measurement) is evidently non-invariant in the most radical way. Not only its translation to a different reference frame is impossible, its very existence is restricted to only one preferred basis. If we were to apply these ideas in the classical case, a rabbit running through a field would have the property of having {\it velocity} (or {\it position}) only for an observer in the platform, but not for an observer in the train ---or anywhere else. Each reference frame would describe a series of different (inconsistent) states of the rabbit.  Thus, the very existence of the rabbit would become untenable. The fact there is no consistent translation of the notion of operational purity, not even in principle, becomes explicit through the derivation of a Corollary to the Kochen-Specker theorem in \cite{deRondeMassri16} which shows that there does not exist a binary valuation for the dynamical properties that constitute a physical system. Indeed, we define as {\it Value Invariance of Dynamical Properties} that the set of dynamical properties that constitute a physical system must be invariant under transformations of frames or coordinates. We say that the context $\mathcal{A}$ \emph{commutes} with the context $\mathcal{B}$ if $AB=BA$ for all $A\in\mathcal{A}$ and $B\in\mathcal{B}$. In particular, if $\mathcal{A}$ is maximal and commutes with $\mathcal{B}$, then $\mathcal{B}\subseteq\mathcal{A}$ and any {\it Local Valuation} defined over $\mathcal{A}$ is defined  over $\mathcal{B}$.
We have the following result,

\begin{theo}\label{teo-VI}
Let $v$ be a {\it Local Binary Valuation} defined over a  maximal context $\mathcal{A}$ and let $x \in {\cal H}$ be any vector. There exists a rotation of $x$ where $v$ is defined and there exists a rotation where $v$ is not defined. In particular, binary valuations are not preserved under rotations.
\end{theo}
\proof see \cite{deRondeMassri16}.
\qed\\

\noindent The previous theorem implies that under a rotation in  ${\cal H}$ the binary valuation is lost. Even though the vector $x \in {\cal H}$ is fixed, the coordinate system is fundamental in order to valuate $x$. We must choose another valuation or else the value of $x$ may not be defined. Consequently, we find the following Corollary to Kochen-Specker Theorem.

\begin{coro}\label{CoroKS}{\bf}
If the dimension of the Hilbert space ${\cal H}$ is greater that $2$, then the Value Invariance of Dynamical Properties of a vector in ${\cal H}$ is precluded. 
\end{coro}

\noindent This Corollary makes explicit the fact that the operational definition of pure state can only make sense in a single basis, and consequently, cannot be regarded as an invariant concept.

On the contrary, mathematical physicists who care not so much about experimental testing of theories and are much more comfortable with abstract definitions apply to their reasonings a different definition of pure state ---one which is reflected in the last part of the just mentioned passage by Nielsen and Chuang. This definition makes reference to a purely abstract mathematical feature of vectors, namely, that when considered as density operators their norm is 1, that is, $\rho$ is a pure state if $\mbox{Tr}(\rho^2)=1$, or equivalently\footnote{A density matrix can be diagonalized, thus giving a set of eigenvalues $0\le\lambda_1\le\ldots<\lambda_n\le 1$ with $\sum_i \lambda_i = 1$. If $\mbox{Tr}(\rho^2)=1$, then $\lambda_1=\ldots=\lambda_{n-1}=0$ and $\lambda_n=1$. Hence, $\mbox{rk}(\rho)=1$ and then  $\rho=|v\rangle\langle v|$ and $\rho=\rho^2$. Conversely, if $\rho=\rho^2$ it has eigenvalues $0$ or $1$, but from $\sum_i \lambda_i = 1$ it follows $\lambda_1=\ldots=\lambda_{n-1}=0$ and $\lambda_n=1$. Hence, $\mbox{Tr}(\rho^2)=1$.} when $\rho=\rho^2$. 
As remarked by Brian Hall in \cite{Hall13}: ``There are several different ways of characterizing the pure states as a subset of the density matrices. First, it is not hard to see that a density matrix $\rho$ is a pure state if and only if $\mbox{Tr}(\rho^2) = 1$. Second, the set of density matrices is a convex set, since if $\rho_1$ and $\rho_2$ are non-negative and have trace 1, then so is 
$\lambda\rho_1+(1-\lambda)\rho_2$, for $0 < \lambda < 1$. [T]he pure states are precisely the extreme points of this set. That is, a density matrix $\rho$ is a pure state if and only if it cannot be expressed as $\rho=\lambda\rho_1+(1-\lambda)\rho_2$ where $\rho_1$ and $\rho_2$ are distinct density matrices and $\lambda$ belongs to $(0,1)$.'' (see \cite[Def. 19.11]{Hall13}).
The latter definition of pure state in terms of convex geometry leads to the development of several mathematical theories and applications to information theory, (see \cite{BZ06} for a modern approach) as well as to philosophical debates about QM (e.g., \cite{AertsSassoli14}). 

The important point is that this abstract mathematical definition makes no reference whatsoever to any specific basis. It is completely independent of reference frames. 
\begin{definition}[Trace-Invariant Purity]\label{pure} An abstract unit vector in Hilbert space $\Psi$ with no reference to a specific basis\footnote{We distinguish here between the purely abstract vector $\Psi$ and its specific representation in a basis $|\psi\rangle$. As shown in detail in \cite[Sect. 4]{deRondeMassri20} this distinction becomes explicitly visualizable through the use of graph theory (see figures 1 and 6 of the mentioned reference).} which understood as a density operator $\rho$ is a projector, namely, where Tr$(\rho^2) = 1$ or $\rho = \rho^2$. 
\end{definition}

\noindent In this case, unlike {\it operational purity}, the notion of pure state is defined in terms of an invariant abstract property characterizing vectors ---as well as families of matrices---, namely, the trace. As discussed in \cite{deRondeMassri18a}, in purely abstract mathematical terms we can always talk about a vector without making reference to a basis. Thus, the abstract vector $\Psi$ makes reference to the state $|\psi\rangle$ but also to any rotation, as for example, $\sum a_i|\varphi\rangle_i$. Now, even though both can be considered as being the same pure state because they have the same trace, while the state $|\psi\rangle$ will be certain if measured, the state $\sum a_i|\varphi\rangle_i$ will be not. If certainty is regarded as a property characterizing the purity of a state, $\sum a_i|\varphi\rangle_i$ cannot be considered as (operationally) pure. While both states will be {\it trace-invariant pure} they will not be {\it operationally pure}. Even more importantly, this second definition of pure state in terms of the trace (or rank) has no operational content whatsoever. This latter definition lacks what Einstein saw as a necessary constituent of a physical concept, namely, the possibility of discovering whether or not it is fulfilled in an actual case.

The difference between these two notions of purity become even more explicit once we recognize that there is no equivalence between them. {\it Trace-invariant purity} does not imply {\it operational purity}. This can be also seen from a mathematical point of view. The following theorem exposes the non-equivalence of these notions. 
\begin{theo}
Let $\rho$ be a density matrix. If $\rho$ is operational pure,
then $\rho$ is trace-invariant pure. The converse does not hold.
\end{theo}
\proof
 While {\it operational purity} translates into ``$\rho=\rho_{op}$, where $\rho_{op}$ is a matrix having a $1$ at position $(1,1)$ and $0$ elsewhere'', {\it trace-invariant purity} is simply $\mbox{Tr}(\rho^2)=1$. Clearly if $\rho=\rho_{op}$,  we have $\mbox{Tr}(\rho^2)=1$, but the converse does not hold. Indeed, if $\mbox{Tr}(\rho^2)=1$ then for any change of basis, the new matrix $U\rho U^\dag$ also satisfies $\mbox{Tr}((U\rho U^\dag)^2)=1$ (here, we see that the definition of trace invariant-purity is context-independent). However, among all bases, there exists only one basis formed by eigenvectors such that the new matrix is diagonal and equal to $\rho_{op}$. Hence, the definition of operational purity depends on a preferred basis. In particular, trace invariant purity does not imply operational purity.\qed\\

In general terms we have that while {\it operational purity} implies {\it trace-invariant purity} the converse is not true: 
{\small 
$$\text{{\it Operational Purity}} \ \ \Rightarrow \ \ \text{{\it Trace-Invariant Purity}}$$ 
$$\ \text{{\it Trace-Invariant Purity}} \ \ \nRightarrow \ \ \text{{\it Operational Purity}}$$
}
This operational-metaphysical conundrum might have its root in a confusion present within the specialized literature between {\it mathematical equivalence} and {\it physical equivalence}. Indeed, as we discussed in section 2, that which is equivalent from a mathematical perspective of analysis is not a necessarily equivalent from a physical standpoint. As explicitly remarked by Einstein, a physical concept does not exist ---for the physicist--- until she has the possibility of discovering whether or not the concept is fulfilled in an actual case. But this is not necessarily provided by mathematical definitions which ---due to their level of abstraction--- may or may not make reference to specific experimental situations. In particular, because the trace does not relate to any basis the link with experience is clearly broken. In order to fix ideas, given a pure state written in the $x$-basis as a unit vector $| \uparrow_x \rangle$, we know that if we measure this state in the same $x$-basis we will obtain with certainty a result related with this state. The probability of observing a measurement related to the state $| \uparrow_x \rangle$, given we prepare the experimental set up in the $x$-basis, will be equal to unity. Of course, in this last statement it is the conditional which is of outmost importance `{\it given} we prepare the experimental set up in the $x$-basis'. However, if we consider the same unit vector now measured in the $z$-basis, for which we represent our state as $ c_1 | \uparrow_z \rangle + c_2 | \downarrow_z \rangle$, there will be no certainty with respect to the measurements that could be performed. While the state written in the $z$-basis is not {\it operationally pure}, both $| \uparrow_x \rangle$ and $ c_1 | \uparrow_z \rangle + c_2 | \downarrow_z \rangle$ have the the same trace and consequently will be {\it trace-invariantly pure}. While from the trace perspective both states can be regarded as being the same, from the operational perspective of purity (or certainty) they cannot. 

\smallskip
 
To conclude, the definitions of purity are not equivalent. The operational definition of purity has only meaning in terms of a preferred basis. Regardless of the difficulties to argue in favor of such basis dependent definition, the mathematical trace definition of pure state is independent of any basis but has no operational counterpart. Or in other words, it is mathematically invariant but not operationally invariant. Thus, the inconsistency becomes evident. Depending on the choice of the definition, the concept of purity is either invariant or non-invariant, operational or purely abstract. This situation is obviously problematic for any physical theory that would attempt to provide a consistent meaning to the concepts it talks about. Against the primacy of preferred basis and pure states over matrices the authors of this paper believe there is a way out of this conundrum by returning to the democratic principle of modern physics according to which all reference frames and states must be considered on equal footing. A first step in this direction has been already given in \cite{deRondeMassri18a, deRondeMassri20} where the definition of state presents an equivalence between all matrices independently of the choice of any basis or their rank. We leave this specific discussion for a future work.

%Arthur Fine: ``Originally, Schr\"odinger thought that his states described a fuzzy bit of reality (something like charge density). But, in correspondence in, Einstein persuaded him that this was not correct, by directing his attention to the case of an unstable pile of gunpowder. After a while, the gunpowder will either explode or not, but the state vector of the whole system at that time would be a superposition involving both exploded and not-exploded component terms. As Einstein wrote, `Through no art of interpretation can this be understood as a description of reality, since there is no intermediary between exploded and not-exploded.' Einstein's gunpowder is the forerunner of Schr\"odinger's cat, whose description Schr\"odinger provided in the next round of their correspondence. These examples show cleanly (no issue of locality here) that if we want to think of the quantum state as describing some bit of reality, then ---as in the case of the gunpowder or the cat--- the quantum description is incomplete. Maximum information, quantum-style, leaves out quite a lot.'' \cite[p. 94]{Schlosshauer11}

\section*{Acknowledgements} 

We want to thank the insightful suggestions of three anonymous referees which allowed us to make the main point of the text substantially more clear. This work was partially supported by the following grants: FWO project G.0405.08 and FWO-research community W0.030.06. CONICET RES. 4541-12, the Project UNAJ 80020170100058UJ and PIO-CONICET-UNAJ 15520150100008CO ``Quantum Superpositions in Quantum Information Processing''. The authors state that there is no conflict of interest.


\begin{thebibliography}{1}

\bibitem{AertsSassoli14} Aerts, D. $\&$ Sassoli di Bianchi, M., 2014, ``The extended Bloch representation of quantum mechanics and the hidden-measurement solution to the measurement problem'', {\it Annals of Physics}, {\bf 351}, 975-1025.

\bibitem{AertsSassoli17} Aerts, D. $\&$ Sassoli di Bianchi, M., 2017, ``Do Spins Have Directions?'', {\it Soft Computing}, {\bf 21}, 1483-1504.

%\bibitem{Born53} Born, M., 1953, ``Physical Reality", {\it Philosophical Quarterly}, {\bf 3}, 139-149.

\bibitem{BeltramettiCassinelli81} Beltrametti, E.G. $\&$ Cassinelli, G., 1981, {\it The Logic of Quantum Mechanics}, Addison-Wesley, Massachusetts.

\bibitem{BZ06} Bengtsson I. $\&$ Zyczkowski, K., 2006, {\it Geometry of Quantum States: An Introduction to Quantum Entanglement}, Cambridge University Press, Cambridge.

\bibitem{Bohr35} Bohr, N., 1935, ``Can Quantum Mechanical Description of Physical Reality be Considered Complete?'', {\it Physical Review}, {\bf 48}, 696-702.

\bibitem{Cartwright72} Cartwright, N., 1972, ``A Dilemma for the Traditional Interpretation of Quantum Mixtures'', {\it Proceedings of the Biennial Meeting of the Philosophy of Science Association}, Vol. 1972, pp. 251-258.

%\bibitem{daCostadeRonde16} da Costa N. $\&$ de Ronde, C., 2016, ``Revisiting the Applicability of Metaphysical Identity in Quantum Mechanics'', preprint. (quant-ph:1609.05361) 

%\bibitem{deRonde16a} de Ronde, C., 2016, ``Probabilistic Knowledge as Objective Knowledge in Quantum Mechanics: Potential Immanent Powers instead of Actual Properties'', in {\it Probing the Meaning of Quantum Mechanics: Superpositions, Semantics, Dynamics and Identity}, pp. 141-178, D. Aerts, C. de Ronde, H. Freytes and R. Giuntini (Eds.), World Scientific, Singapore.

%\bibitem{deRonde18a} de Ronde, C., 2018, ``Quantum Superpositions and the Representation of Physical Reality Beyond Measurement Outcomes and Mathematical Structures'', {\it Foundations of Science}, {\bf 23}, 621-648.

%\bibitem{deRondeFM18} de Ronde, C. $\&$ Fern\'adez-Mouj\'an, R., 2018, ``Epistemological vs. Ontological Relationalism in Quantum Mechanics: Relativism or Realism?'', in {\it Probing the Meaning of Quantum Mechanics. Superpositions, Semantics, Dynamics and Identity}, pp. 277-317, D. Aerts, M.L. Dalla Chiara, C. de Ronde and D. Krause (Eds.), World Scientific, Singapore. 

%\bibitem{deRondeFreytesSergioli19} de Ronde, C., Freytes, H. $\&$ Sergioli, G., 2019, ``Quantum probability: a reliable tool for an agent or a reliable source of reality?'', {\it Synthese}, DOI: 10.1007/s11229-019-02177-x.

\bibitem{deRondeMassri16} de Ronde, C. $\&$ Massri, C., 2017, ``Kochen-Specker Theorem, Physical Invariance and Quantum Individuality'',  {\it Cadernos de Hist\'oria e Filosofia da Ci\^encia}, {\bf 2}, 107-130.

\bibitem{deRondeMassri18a} de Ronde, C. $\&$ Massri, C., 2018, ``The Logos Categorical Approach to Quantum Mechanics: I. Kochen-Specker Contextuality and Global Intensive Valuations.'', {\it International Journal of Theoretical Physics}, DOI: 10.1007/s10773-018-3914-0. 

%\bibitem{deRondeMassri18b} de Ronde, C. $\&$ Massri, C., 2019, ``The Logos Categorical Approach to Quantum Mechanics: II. Quantum Superpositions and Intensive Values'', {\it International Journal of Theoretical Physics}, {\bf 58}, 1968-1988.

%\bibitem{deRondeMassri19} de Ronde, C. $\&$ Massri, C., 2019, ``A New Objective Definition of Quantum Entanglement as Potential Coding of Intensive and Effective Relations.'', {\it Synthese}, DOI: 10.1007/s11229-019-02482-5. 

\bibitem{deRondeMassri20} de Ronde, C. $\&$ Massri, C., 2020, ``Beyond Purity and Mixtures in Categorical Quantum Mechanics'', preprint, sent. (quant-ph:2002.04423)

\bibitem{d'Espagnat76} D'Espagnat, B., 1976, {\it Conceptual Foundations of Quantum Mechanics}, Benjamin, Reading MA.

\bibitem{Dieks10} Dieks, D., 2010, ``Quantum Mechanics, Chance and Modality'', {\it Philosophica}, {\bf 83}, 117-137.

\bibitem{Dirac74} Dirac, P.A.M., 1974, {\it The Principles of Quantum Mechanics}, 4th Edition, Oxford University Press, London.

\bibitem{Einstein20} Einstein, A., 1920, {\it Relativity. The Special and General Theory}, Henry Holt $\&$ Company, New York. 

\bibitem{EPR} Einstein, A., Podolsky, B. $\&$ Rosen, N., 1935, ``Can Quantum-Mechanical Description be Considered Complete?'', {\it Physical Review}, {\bf 47}, 777-780.

\bibitem{FLSG13} Freytes, H., Ledda, A., Sergioli, G. $\&$ Giuntini, R., 2013, ``Probabilistic logics in quantum computation'' in {\it New Challenges to Philosophy of Science}, pp. 49-57, H. Andersen, D. Dieks, W.J. Gonzalez, Th. Uebel and G. Wheeler, Springer, Berlin.

\bibitem{Hall13} Hall, B.C., 2013, {\it Quantum Theory for Mathematicians}, Springer, Berlin. 

\bibitem{Jauch68} Jauch, J.M., 1968, {\it Foundations of Quantum Mechanics}, Addison-Wesley, Massachusetts.

\bibitem{KS} Kochen, S. $\&$ Specker, E., 1967, ``On the problem of Hidden Variables in Quantum Mechanics'', {\it Journal of Mathematics and Mechanics}, {\bf 17}, 59-87. Reprinted in Hooker, 1975, 293-328.

\bibitem{Mermin98b} Mermin, D., 1998, ``What is Quantum Mechanics Trying to Tell Us?", {\it American Journal of Physics}, {\bf 66}, 753-767.

\bibitem{Mittelstaedt78} Mittelstaedt, P., 1978, {\it Quantum Logic} D. Reidel Publishing Company, Berlin.

\bibitem{NielsenChuang00} Nielsen, M. $\&$ Chuang, I., 2000, {\it Quantum Computation and Quantum Information}, Cambridge University Press. 

\bibitem{Peres02} Peres, A., 2002, {\it Quantum Theory: Concepts and Methods}, Kluwer Academic Publishers, Dordrecht. 

\bibitem{Piron76} Piron, C. 1976, {\it Foundations of Quantum Physics} W.A. Benjamin Inc., Reading, New York. 

\bibitem{Schlosshauer11} Schlosshauer, M. (Ed.), 2011, {\it Elegance and Enigma. The Quantum Interviews}, Springer-Verlag, Berlin.

\bibitem{Scholz18} Scholz, E., 2018, ``Weyl's search for a difference between `physical' and `mathematical' automorphisms'.'' {\it Studies in History and Philosophy of Modern Physics}, {\bf 61}, 57-67.

\bibitem{Smets05} Smets, S., 2005,  ``The Modes of Physical Properties in the Logical Foundations of Physics'', {\it Logic and Logical Philosophy}, {\bf 14}, 37-53.

\bibitem{Timpson10} Timpson, C., 2010, ``Information, Immaterialism, Instrumentalism: Old and New in Quantum Information'', in {\it Philosophy of Quantum Information and Entanglement}, pp. 208-228, Bokulich and G. Jaeger (Eds.), Cambridge University Press, Cambridge. 

\bibitem{VF80} Van Fraassen, B.C., 1980, {\it The Scientific Image}, Clarendon, Oxford.

\bibitem{Weyl38} Weyl, H, 1938, {\it The Classical Groups},  Princeton University Press, Princeton.

\bibitem{Weyl52} Weyl, H, 1952, {\it Symmetry}, Princeton University Press, Princeton.

%\bibitem{VN32} Von Neumann, J., 1955, {\it Mathematical Foundations of Quantum Mechanics.} Trans. Robert T. Geyer. Princeton, Princeton University Press, Princeton.



\end{thebibliography}
\end{document}